\documentstyle[twoside,fleqn,espcrc2]{article}

\input epsf

\def\BNL {Department of Physics, Brookhaven National Laboratory, Upton, NY 11973}
\def\WU {Department of Physics, Washington University, St. Louis, M0 63130}

\newcommand{\BE}{\begin{equation}}
\newcommand{\EE}{\end{equation}}
\newcommand{\BEA}{\begin{eqnarray}}
\newcommand{\EEA}{\end{eqnarray}}

\newcommand{\gbeta}{6/g^2}

\newcommand{\kl}{\kappa_{l}}
\newcommand{\kh}{\kappa_{h}}
\newcommand{\Bd}{B^0}
\newcommand{\Bs}{B^0_s}
\newcommand{\MB}{M_{bd}}
\newcommand{\MS}{M_{bs}}
\newcommand{\MP}{M_{hl}}
\newcommand{\MC}{M_{hc}}
\newcommand{\MHS}{M_{hs}}

\begin{document}

\title{ SU(3) Flavor Breaking in Hadronic Matrix Elements for $B - \bar B$
Oscillations } 
\author{C. Bernard\address{ \WU }, T. Blum\address{\BNL}, and A. Soni$\,\null^{\rm b}$ }

\begin{abstract}
We present an analysis, using quenched configurations at $6/g^2=$5.7, 5.85, 6.0, 
and 6.3 of the matrix element 
$\MP\equiv\langle \bar P_{hl}|\bar h \gamma_\mu
(1-\gamma_5)l \bar h \gamma_\mu(1-\gamma_5)l|P_{hl}\rangle$ 
for heavy-light pseudoscalar mesons. The results are extrapolated to the
physical $B$ meson states, $\Bd$ and $\Bs$. 
We directly compute the ratio $\MS/\MB$, and obtain the preliminary result
$\MS/\MB=1.54(13)(32)$.
A precise value of this SU(3) breaking ratio is important
for determining $V_{td}$ once the mixing parameter $x_s$ 
for $\Bs-\bar\Bs$ is measured experimentally. We also determine values
for the corresponding B parameters, $B_{bs}(2 \rm{GeV})=B_{bd}(2 \rm{GeV})=1.02(13)$, which we
cannot distinguish in the present analysis.
\end{abstract}

\maketitle

\section{ Introduction }
We present a direct calculation of the $\Delta F=2$ heavy-light
mixing matrix element,
\BEA
\label{matrix hl}
\MP(\mu) \equiv \langle \bar P_{hl}|\bar h \gamma_\rho
(1-\gamma_5)l \bar h \gamma_\rho(1-\gamma_5)l|P_{hl}\rangle,\!\! 
\EEA
where we have suppressed the scale dependence of the local four quark operator.
As is well known, these matrix elements govern $\Bd-\bar\Bd$ and $\Bs-\bar\Bs$
oscillations~\cite{FLYNN,SONI}. In the above $h$ and $l$ denote heavy and light quark
fields, $P_{hl}$ the corresponding pseudoscalar meson, and $\mu$ is the energy 
scale appropriate to the calculation. In particular, we present a first direct 
calculation
of the SU(3) flavor breaking ratio,
\BEA
\label{ratio}
r_{sd}=\MS(\mu)/\MB(\mu)
\EEA

Our preliminary result is $r_{sd}=1.54(13)(32)$ where the first error is
statistical and the second is from uncertainty in extrapolating 
to the $B$ mass and 
to $a\to 0$.
The importance of this ratio is that, in conjunction with the experimental measurement
of $\Bs-\bar\Bs$ oscillation (when that becomes available), it should allow the 
cleanest extraction of the crucial CKM parameter $V_{td}$. 
We note that the above value of $r_{sd}$
is lower than the result reported at the conference (1.81) which is due primarily to
a change in how we extrapolate to the $B$; the difference gives a systematic error.

Presently, $V_{td}$ is deduced from $\Bd-\bar\Bd$ oscillation via\cite{BURAS}  
\BEA
\label{mixing par}
x_{bd}=\frac{(\Delta M)_{bd}}{\Gamma_{bd}}\propto m_{bd}^2
B_{bd}(\mu)f_{bd}^2 |V_{td}|^2
\EEA
where $m_{bd}$, $\Gamma^{-1}_{bd}\equiv\tau_{bd}$, and $f_{bd}$ 
are the mass, life time, and
decay constant of the $\Bd$ meson, and $(\Delta M)_{bd}$ is the mass difference of the
two mass eigenstates of the $\Bd-\bar\Bd$ system. $x_{bd}$ is the mixing parameter 
characterizing the oscillation and has been determined experimentally, 
$x_{bd}=0.71(6)$~\cite{PDB}. $B_{bd}$ is the so called bag, or B, parameter.
To extract $V_{td}$ from Eq.(~\ref{mixing par}) 
requires knowledge of two hadronic
matrix elements, $f_{bd}$ and $B_{bd}$. 
These are being calculated by using lattice and other methods.
$f_{bd}$ may eventually even be measured experimentally through,
for example, the decay $B\to \tau \nu_\tau$. 
However, $B_{bd}$ is a purely theoretical
construct which is inaccessible to experiment. 
Thus determination of $V_{td}$ from experiment through use of Eq.~\ref{mixing par}
will ultimately be limited by the precision of the
nonperturbative quantity
$f^2_{bd}B_{bd}$. These parameters are related to the matrix element,
Eq.~(\ref{matrix hl}), via
\BEA
\label{B param}
\MB(\mu)=\frac{8}{3}f_{bd}^2 m_{bd}^2 B_{bd}.
\EEA

Now making the replacement $d\to s$ in Eq.~(\ref{mixing par}) and taking the
ratio with Eq.~(\ref{mixing par}), we arrive at an alternate way to 
extract $V_{td}$, 
\BEA
\label{ckm ratio}
\frac{|V_{td}|^2}{|V_{ts}|^2}&=&r_{sd} 
\frac{\tau_{bs}}{\tau_{bd}} 
\frac{x_{bd}}{x_{bs}}
\EEA
Note that $V_{ts}$ in Eq.~(\ref{ckm ratio}) is related
by three generation unitarity to $V_{cb}$ and is therefore already quite
well determined, $|V_{ts}|\approx|V_{cb}|=0.041 \pm 0.003\cite{PDB}$.
The important distinction between using Eq.~(\ref{ckm ratio}) instead of 
Eq.~(\ref{mixing par}) is that the former requires only 
knowledge of {\it corrections} to SU(3) flavor symmetry while the latter 
requires the {\it absolute} value of the matrix element $\MB$.
It is also important to realize that since $r_{sd}$ is a ratio of two 
very similar hadronic
matrix elements, it is less susceptible to common systematic 
errors in lattice calculations, including scale dependence, 
matching of continuum and lattice
operators, and heavy quark mass dependence. Indeed, the ratio
$r_{sd}$ is, to an excellent approximation, renormalization group
invariant, even though the individual matrix elements
$M_{bs}$ and $M_{bd}$ are scale dependent.

\section{ Simulations and Results}
Table~(\ref{lat table}) summarizes our quenched lattices and
the valence Wilson quark hopping parameters, $\kl$ and $\kh$,
used to construct quark propagators. For each $\kl$ and 
$\kh$ in Table~\ref{lat table} we calculate a quark propagator 
using a single point source at the center of the lattice and
a point sink.

\begin{table*}[hbt]
\caption{Summary of simulation parameters.}
\label{lat table}
\begin{tabular}{ccccc}
\hline
$\gbeta$ & conf. & size & $\kl$ &$\kh$ \\
\hline
5.7&	83&$	16^3\times 33$&	0.160 0.164 0.166&0.115 0.125 0.135 0.145\\
5.85&	100&$	20^3\times 61$&	0.158 0.159 0.160&0.107 0.122 0.130 0.138 0.143\\
6.0&	60&$	16^3\times 39$&	0.152 0.154 0.155&0.103 0.118 0.130 0.135 0.142\\
6.0&	100&$	24^3\times 39$&	0.152 0.154 0.155&0.103 0.118 0.130 0.135 0.142\\
6.3&	60&$	24^3\times 61$&	0.148 0.149 0.150 0.1507&0.100 0.110 0.125 0.133 0.140\\
\hline
\end{tabular}
\end{table*}

In Fig.~\ref{mll b63} we show example results at $\gbeta=6.3$ for $\MP$
versus $\kh$ for each value of $\kl$. The lattice matrix elements are found
through simultaneous fits to the three point function corresponding to
Eq.~\ref{matrix hl}
and the two point function of the corresponding heavy-light interpolating operator.
In Fig.~\ref{mll b63} and the following, we have already matched the lattice 
operator to the continuum one by
using the one loop perturbative result from Refs.~\cite{MART,BDS}.
The scale dependent renormalization $Z$ factor is calculated in the NDR 
scheme at scale $\mu=2.0$ GeV.
The coupling is tadpole improved and evaluated at scale $1/a$,
and we include the KLM normalization for heavy quarks.

Results for the physical $B$ and $B_s$ meson systems follow from 
a series of fits to the lattice data which we use to extrapolate in the
two parameters $\kh$ and $\kl$. Since the data are highly correlated,
we use covariant fits and a jackknife procedure at each step to account for 
the correlations. We take the form of our fits from chiral perturbation 
theory and expectations based on heavy quark effective theory (HQET).
\begin{figure}[hbt]
    \vspace{-0.2in}
    \vbox{
        \hskip-.2in\epsfxsize=3.0in \epsfbox[0 0 4096 4096]{ll_corr.ps}
    }
    \vskip -0.5in
    \caption{ The matrix element $\MP$ at $\gbeta=6.3$.}
    \label{mll b63}
    \vspace{-0.2in}
\end{figure}

$\kappa_c(\gbeta)$ and $\kappa_s(\gbeta)$ are determined
from either linear or quadratic fits to $m_{ll^\prime}^2$ as a function
of $\kl^{-1}$ and $\kappa_{l^\prime}^{-1}$
($l^\prime$ refers to the strange quark). 
The values for $\kappa_c$ and $\kappa_s$ are summarized in 
Table~\ref{kctable}. Finding $\kappa_s$ requires 
the scale $a$, which we set from $af_\pi$, to determine the lattice value of the kaon mass $a m_K$
($a^{-1}$ is tabulated in Table~(\ref{kctable})).
We note that at $\gbeta=5.7$ the choice of the coupling constant scale
for $Z_A$, the axial current renormalization, has a significant effect on the lattice
spacing determination; $Z_A$ differs by $\sim 7\%$ when the scale changes from
$1/a$ to $\pi/a$.

Next, using chiral perturbation theory for heavy-light mesons~\cite{BOOTH,SY},
we extrapolate $\MP$ to $\kl=\kappa_c$. 
We do not include chiral logarithms in our fits
since the light quark masses used in the extrapolations are relatively heavy, 
$\kl\approx\kappa_s$. The results for $\MP$
at $\gbeta=6.3$ (see Fig.~\ref{mllc}) show a smooth linear behavior. 
The confidence levels for the extrapolations are much lower for $\gbeta=6.0(24^3)$ than 
for the other points; in addition, the effective values of $\MP$ show poor plateaus,
so we exclude this point in our final determination of $r_{sd}$.
\begin{figure}[hbt]
    \vspace{-0.2in}
    \vbox{
        \hskip-.2in\epsfxsize=3.0in \epsfbox[0 0 4096 4096]{mllc.ps}
    }
    \vskip -0.5in
    \caption{\label{mllc} $\MP$ at $\gbeta=6.3$ extrapolated to $\kl=\kappa_c$. Curves
		for the other values of $\gbeta$ are similar.}
    \vspace{-0.2in}
\end{figure}

\begin{table}[hbt]
\caption{Inverse lattice spacing and critical and strange hopping parameters.}
\label{kctable}
\begin{tabular}{cccc}
\hline
$\gbeta$&  $a^{-1}$(GeV)&$\kappa_c$&$\kappa_s$ \\
\hline
5.7&	1.45(10)&	0.16969(10)&	0.1642(7) \\
5.85&	1.64(20)&	0.16163(9) &	0.1577(9)\\
6.0&    2.06(17)&	0.15711(7) &	0.1548(4)\\
6.0&   	2.08(13)&	0.15714(4) &	0.1543(4)\\
6.3&    3.37(47)&	0.15226(16)&	0.1506(4)\\
\hline
\end{tabular}
\vspace{-0.2in}
\end{table}

\begin{figure}[hbt]
    \vbox{
       \hskip-.2in \epsfxsize=3.0in \epsfbox[0 0 4096 4096]{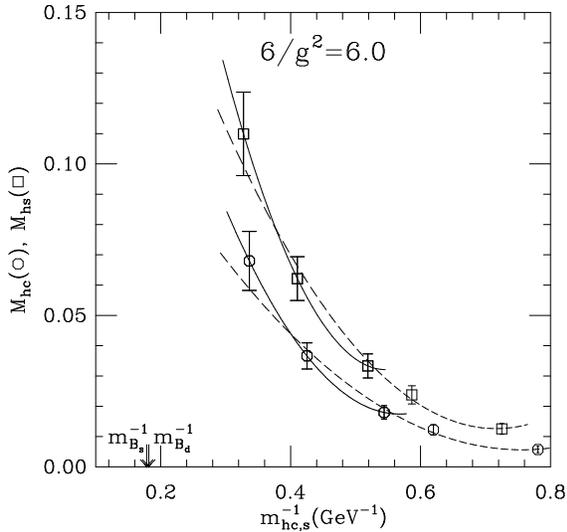}
    }
    \vskip -0.5in
    \caption{\label{Mvsminv} $\MC$ (octagons) and $\MHS$ (squares) as a 
		function of the inverse
		heavy-down(strange) meson mass, at $\gbeta=6.0$. The
		dashed line shows the effect of the lightest points on the fit.}
    \vspace{-0.25in}
\end{figure}

\begin{figure}[hbt]
    \vbox{
       \hskip-.2in \epsfxsize=3.0in \epsfbox[0 0 4096 4096]{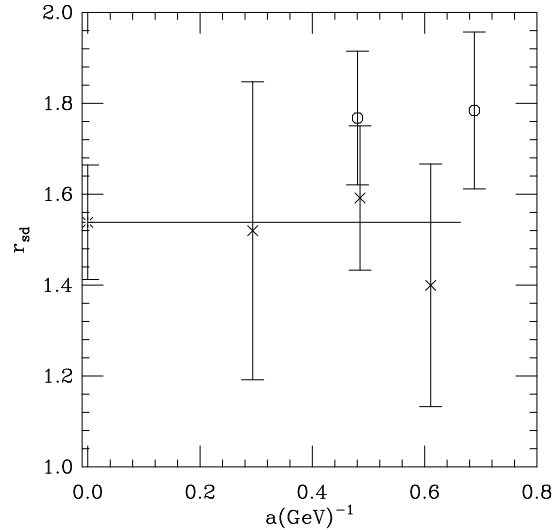}
    }
    \vskip -0.5in
	\caption{\label{Mratio} The SU(3) flavor breaking ratio $\MS/\MB$
		versus the lattice spacing $a$. The points denoted by crosses
		were used in the fit (solid line). The circles ($\beta=5.7$ and
                $\beta=6.0, 24^3$) were omitted, for reasons explained in the
		text.  The burst shows $r_{sd}$ extrapolated to $a=0$.
}
    \vspace{-0.25in}
\end{figure}

Inspired by HQET, we continue by fitting $\MC$ to a polynomial in 
the inverse heavy meson mass, $m_{hc}^{-1}$, and then extrapolating to the $B$ meson 
mass (see Fig.~\ref{Mvsminv}).
Again, we use $f_\pi$ to set the scale. Of course, the same procedure 
can be carried through for the heavy-strange mesons by first extrapolating the
data to $\kappa_s$ instead of $\kappa_c$. The resulting curve is also shown in Fig.~\ref{Mvsminv}. 
Fits which include all mass values (dashed line) generally have low confidence
levels. Moreover, the lighter points  have smaller statistical errors and dominate
the fits, yet they are far from the $B$ mass. We therefore omit the
lightest two points at each value of $\gbeta$ and use a completely constrained fit to extrapolate
to the $B$.  Compared with fitting all points, this results in systematically
lower values of $r_{sd}$, changing the central value from 1.81 (which was reported at the conference)
to 1.54.  We take the 0.27 difference as the systematic error of the extrapolation.
The ratio $\MS/\MB$ is shown as a function of $a(\gbeta)$
in Fig.~\ref{Mratio}. 
At $\gbeta=5.7$, our heaviest mass is still quite far from the
$B$ mass, so we also ignore this point in the extrapolation to the continuum limit.

Generally, we expect the Wilson quark action to introduce
discretization errors of order $a$ in
all observables. However, in a ratio of two similar 
quantities, we might expect a large cancellation of the lowest order discretization
errors. A constant fit gives $\MS/\MB= 1.54(13)$ while a linear fit gives 1.72(67).
Since the coefficient of the linear term is only 0.3 sigma from 0,
we quote the constant fit
result as our central value and use the difference as an estimate of the systematic
error 
of the continuum extrapolation. Adding that error in quadrature with the
systematic error from the extrapolation to the $B$ mass, we get $r_{sd}=1.54(13)(32)$. 

The extraction of the individual values of $\MB$ and $\MS$ is clearly expected
to have larger errors. Conventionally~\cite{CB88,FLYNN,SONI} these matrix
elements are given in terms of the corresponding B parameter defined in Eq.~(\ref{B param}).
Carrying out the above fitting procedure
for $B_{bd}(\mu)$, we find a constant fit yields $B_{bd}(2 \,{\rm GeV})=0.97(3)$ while 
linear extrapolation gives 1.02(13). We cannot however distinguish $B_{bs}(2 \,{\rm GeV})$ 
from $B_{bd}(2 \,{\rm GeV})$ since our data for $B_{hl}$ versus $\kl$ are fit equally well 
to constant or linear forms. This was not true for $\MP$ as is evident from Fig.~\ref{mllc}.
Taking the linear result, we quote $B_{bd}(2 \,{\rm GeV})=B_{bs}(2 \,{\rm GeV})=1.02(13)$.

We recall that until now~\cite{FLYNN,SONI}, lattice results for the SU(3) breaking
ratio $r_{sd}$ have been obtained by using Eqs.~(\ref{ratio}) and (\ref{B param})
and the lattice measurements of $f_{bd(s)}$ and $B_{bd(s)}$.
A reasonable estimate is $f_{bs}/f_{bd}\approx1.13\pm .10$~\cite{FLYNN,SONI,BLS,UKQCD}
(we are presently calculating this ratio on our lattices).
As indicated above, the ratio of B parameters is consistent with unity, and
the ratio of masses is 1.017~\cite{PDB}.
Therefore, the conventional method leads to $r_{sd}\approx 1.32\pm .23$
compared to $1.54\pm .13\pm .32$ obtained with our direct method. 
Thus the two methods are quite consistent.
However, as we have emphasized, the direct method offers many distinct 
advantages, and future lattice computations should be able to improve the precision in
our determination of the 
ratio $r_{sd}$. 

This research was supported by US DOE grants 2FG02-91ER4D628 and DE-AC0276CH0016.
The numerical computations were carried out at the NERSC supercomputing center.

\end{document}